\documentclass[aps,prl,showpacs,twocolumn, superscriptaddress,groupedaddress,reprint]{revtex4}

\usepackage{subfigure}
\usepackage{wrapfig}
\usepackage{textcomp}
\usepackage{graphicx}
\usepackage{amssymb}
\usepackage{amsmath}
\usepackage{bm}
\usepackage{amsmath, amsthm, amssymb}
\usepackage{hyperref}
\usepackage{color}
\usepackage{rotating}

\usepackage{amsfonts}
\usepackage{dcolumn}
\usepackage{float}
\usepackage{bm}

\begin{document}

\author{A. J. White}
\author{L. A. Collins}
\affiliation{Theoretical Division, Los Alamos National Laboratory, Los Alamos, 87545 NM}
\date{\today}

\title{A Fast and Universal Kohn Sham Density Functional Theory Algorithm for Warm Dense Matter to Hot Dense Plasma}

\begin{abstract}
Understanding many processes, e.g. fusion experiments, planetary interiors and dwarf stars, depends strongly on microscopic physics modeling of warm dense matter (WDM) and hot dense plasma. This complex state of matter consists of a transient mixture of degenerate and nearly-free electrons, molecules, and ions. This regime challenges both experiment and analytical modeling, necessitating predictive \emph{ab initio} atomistic computation, typically based on quantum mechanical Kohn-Sham Density Functional Theory (KS-DFT).  However, cubic computational scaling with temperature and system size prohibits the use of DFT through much of the WDM regime.  A recently-developed stochastic approach to KS-DFT can be used at high temperatures, with the exact same accuracy as the deterministic approach, but the stochastic error can converge slowly and it remains expensive for intermediate temperatures (<50 eV).  We have developed a universal mixed stochastic-deterministic algorithm for DFT at any temperature. This approach leverages the physics of KS-DFT to seamlessly integrate the best aspects of these different approaches. We demonstrate that this method significantly accelerated self-consistent field calculations for temperatures from 3 to 50 eV, while producing stable molecular dynamics and accurate diffusion coefficients.
\end{abstract}

\maketitle

The intermittent nature of warm dense matter (WDM)  and hot dense plasmas (HDP) makes calculation of microscopic properties particularly difficult \cite{Graziani_14}. The quantum mechanical interaction of electrons through fermionic antisymmetry, i.e. Pauli exclusion or electron degeneracy, is comparable to the thermal energy. Both the electrostatic interactions and kinetic energies of ions are non-negligible; while the question of determining ``partial charges" on atoms, and the existence of spatial and temporal correlations between ions or even transient molecular structures, creates additional difficulty. However modeling these regimes is critical to a range of systems including planetary bodies\cite{Benuzzi_Mounaix_2014,French_2019,REDMER2011798,Guillot_2005,Guillot:1999aa,Nguyen_2004,Pozzo_2012,Stevenson_2009,Nettelmann_2008,Ehrenreich_2020}, brown and white dwarf stars\cite{10.2307/2892796,Chabrier_1993,Saumon_1992}, inertial fusion energy \cite{Hinkel_2013}, and high-intensity high-energy laser pulse experiments\cite{Falk_2018}. {\color{blue}}

This necessitates an atomistic, \emph{ab initio}, and quantum-mechanical calculation of microscopic properties. For the cooler side of WDM, Kohn-Sham Density Functional Theory (KS-DFT) \cite{Kohn:1965aa} is the gold standard for calculating material properties \cite{Graziani_14,DOE_16}. However, due to the poor computational scaling (both in time and memory) of KS-DFT with temperature, its use is limited to systems typically $\tilde<\,10$ eV. For high temperatures, more approximate orbital-free (OF-) DFT \cite{White:2017aa,PhysRevE.95.043210,White:2019aa,Ticknor:2016aa}  or Path-Intergal Monte Carlo (PIMC) \cite{Driver:2012aa,Driver:2018aa,Dornheim:2019aa}, typically within the fixed node approximation , become the standard . The cost of PIMC calculations significantly increases for lower temperatures. Combining these approximations, and KS-DFT, for different parameter spaces has been successful in generating equation of state data \cite{Militzer:2001aa,Benedict:2014aa}. However, extension into time-dependent quantities, $\emph i.e.$ electronic response properties, stopping-power, absorption spectra, conductivities \emph{etc.}, is not possible with PIMC, and highly approximate for OF-DFT\cite{White:2018aa}.  

Recently, a stochastic alternative (sDFT) to the traditional  deterministic KS-DFT (dDFT) algorithm has been developed \cite{doi:10.1002/wcms.1412,Cytter:2018aa}. The computational costs and memory scale linearly with system size, $V$; the method is nearly trivial to parallelize and converges to the exact KS-DFT result \cite{Baer:2013aa}. Moreover, the computational cost scales inversely to the temperature, \emph{i.e.} $\propto V/T$ rather than cubically as in the traditional algorithm \cite{Cytter:2018aa}, \emph{i.e.}  $\propto V^3T^3$. For moderate WDM to HDP temperatures, roughly from $~5$ to $40$ eV but dependent on system size and density, neither the sDFT nor dDFT are sufficiently efficient for long time molecular dynamics simulations (compared to near-zero temperature dDFT) with reasonable computational resources.  For volume-averaged quantities, error cancelation leads to sub-linear scaling with system size, and $<1\%$ errors in energies can be achieved with  $\approx100$ ``stochastic orbitals" \cite{Baer:2013aa}.  Convergence of the electron density or ion forces is more challenging, especially for ionized systems.\cite{Neuhauser_2014,Li_2019,Chen_2019,Chen_2019_2} 

 We present a universal mixed stochastic - deterministic approach (mDFT), which combines the best aspects of the dDFT and sDFT algorithms and leverages the physics of the KS-DFT problem, to achieve faster and more precise self-consistent field calculations at high temperature. The approach is based on separation of the eigenspectrum of the KS-DFT Hamiltonian into a low-energy deterministic and high-energy stochastic segments. Like sDFT, converged mDFT provides the exact same accuracy as converged dDFT calcualtions, but mDFT can be much more precise than sDFT for a given computational time (or faster for a given precision). Since the sDFT and dDFT are different linear-algebra techniques solving the same KS-DFT problem, this generalization is pointedly simpler and more robust than merging KS-DFT with a more approximate method, such as OF-DFT or plane-wave electrons \cite{Zhang_2016}. For example, high energy electrons are still properly orthogonal to low energy states, preventing overlap with core regions. 

We will first describe the computational details, pertinent to computational time and scaling, of the dDFT and sDFT approaches. Then, we will detail the algorithm for mDFT. We will then present tests for the accuracy, precision, and computational timing of the mDFT algorithm with at a range of temperatures, from 1 to 50eV.  

In  KS-DFT the  effective single-particle DFT Hamiltonian is defined as (all equations in atomic units),
\begin{align}
\label{ham}
{\hat H}_{DFT} = -\nabla^2/2 +\int d{\vec r}^{\,3} \frac{\rho({\vec r'})}{\vert {\vec r}- {\vec r'} \vert} + {\hat V}_{xc} +  {\hat V}_{ext},
\end{align}
where $\nabla$ is the gradient operator, $\rho$ is the electronic density, ${\hat V}_{xc}$ is the exchange-correlation potential, and ${\hat V}_{ext}$ is the external potential due to the ions or electric fields. The latter potentials may be local potentials or non-local operators. Within the finite-temperature Mermin-Kohn-Sham theory\cite{Mermin:1965aa}, the equilibrium electron density is given by 
 \begin{align}
\label{Trace}
\rho(r)= \text{Tr}\{f(\hat H_{DFT},\mu,T)\}, \text {with}
\\
\nonumber
f({\hat H}_{DFT},\mu,T)={2\over 1-e^{({\hat H}_{DFT}-\mu)/T}},
\end{align}
where $\mu$ is the electron chemical potential, $T$ is the temperature, and $f$ is the Fermi-Dirac distribution function. In dDFT the trace is taken over the orbitals, $\psi$s (eigenvectors of $\hat H_{DFT}$), whereas in sDFT the trace is a taken over stochastic orbitals, $\chi$s \cite{doi:10.1080/03610919008812866}. 

  \begin{figure}[t]
\begin{center}
\includegraphics[width=0.75\columnwidth] {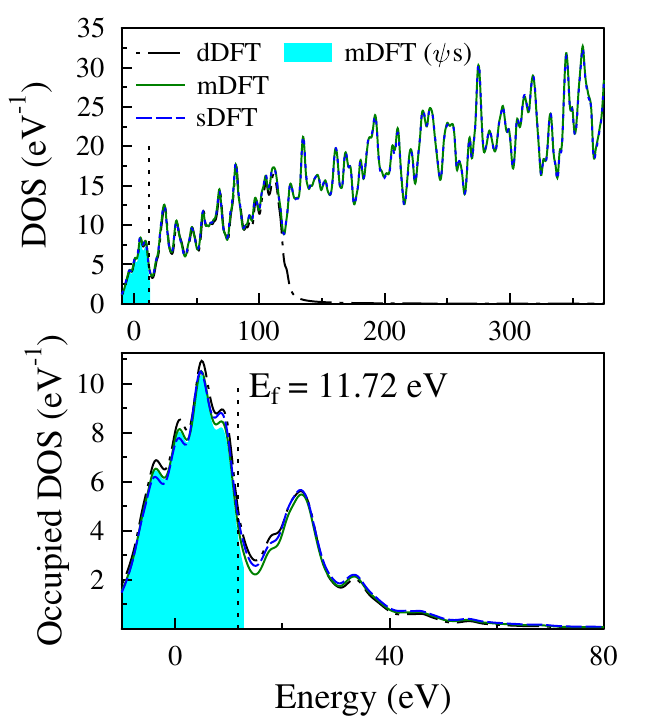}
\caption{(Color online) (Top) Density of states (DOS) of 64 atoms diamond structure carbon (3.51 g/cc) at 10 eV. See text for explicit form. dDFT (dash-dot line), sDFT( dashed line), mDFT(solid line), and mdFT deterministic component (solid fill under) are shown. (Bottom) The occupied DOS. Individual states are represented as Lorentzians with a 2 eV broadening. Fermi energy, Ef, shown as dashed verticel line.     }
\label{dos}
\end{center}
\end{figure}

For dDFT, the computational cost is dominated by eigendecomposition, \emph{i.e.} solving ${\hat H}_{DFT}\, \psi_a = \varepsilon_a \psi_a$. Solving for $\psi$'s can be efficiently achieved for $N_{\psi}\to O(100)$, but for WDM to HDP temperatures $N_{\psi}\to O(1000)$, with $N_{\psi} \propto T^{3/2}$, the need to orthogonalize all orbitals leads to the $T^3$ scaling \cite{Cytter:2018aa}. The lowest energy orbitals carry the largest contribution to the density (due to $f$), but at higher energies, the density of states grows. For high temperatures this leads to many orbitals with small $f$, such that collectively they are important, but they are increasingly costly to resolve individually. 

For sDFT, the computational cost  is dominated by finding the ``occupied" stochastic orbitals $X_b=\sqrt{f({\hat H}_{DFT},\mu,T)} \chi_b$. This involves the Chebychev expansion of $\sqrt{f({\hat H}_{DFT},\mu,T)}$, which converges $\propto E_{cut}/T$ \cite{Cytter:2018aa}. $E_{cut}$ is the cut-off energy for the plane-wave basis. The dominant operation is then the application of ${\hat H}_{DFT}$ to an intermediate vector, $  \phi\to{\hat H}_{DFT}\phi$, which is also a significant operation in dDFT. For periodic plane-wave systems considered here, this operation scales nearly-linear as $N_g \text{Ln}(N_g)$ due to fast-Fourier transforms (FFT), where $N_g$ is the number of plane-waves($\propto$ the volume and $E_{cut}^{1/2}$). For moderate WDM temperatures, the Chebychev expansion still requires many terms \cite{Cytter:2018aa}. The stochastic orbital is given by $\chi_b = 1/\sqrt{N_\chi d{\vec r}^{\,3}}e^{i2\pi \theta_b({\vec r})}$, where $\theta_b({\vec r})$ is an independent random number between $0$ and $1$. It can naturally be expressed as a linear combination of $\psi$s: $\chi_b({\vec r})=\sum_a^\infty c_{b,a}({\vec \theta}_b)\,\psi_a({\vec r})$. 

The mDFT generalization of the dDFT and sDFT approaches can be realized by:
\newline

\noindent 1)  Generate and store $\chi$s (or random number seeds). 

\noindent 2) Begin self-consistent field (SCF) loop: Generate potentials in Eq.\ref{ham} from trial $\rho$.

\noindent 3) Efficiently solve for a reduced set of $\psi$s using an appropriate eigensolver (conjugate gradient\cite{KRESSE199615}, LOBPCG\cite{doi:10.1137/S1064827500366124}, Chebychev Filtering\cite{LEVITT201598}).

\noindent 4) Project the $\psi$s out of the $\chi$s to form $\tilde \chi$s:
\begin{align}
\tilde\chi_b=\chi_b -\sum^{N_\psi}_a c_{b,a}\psi_a.
\end{align}
\noindent 5) Find the ``occupied" $\tilde\chi s$: $\tilde X_b=\sqrt{f({\hat H}_{DFT},\mu,T)} \tilde \chi_b$. 

\noindent 6)  Solve for the new density and iterate until SCF loop ($\mu$, $\varepsilon$, $\rho$) converges: 
\begin{align}
{\rho({\vec r})}= \sum_a^{N_\psi} f(\varepsilon_a,\mu,T)  \vert \psi_a({\vec r})\vert^2   +  \sum_b^{N_\chi} \vert \tilde X_b({\vec r}) \vert^2 
\end{align}
\noindent 7) Use the mixed density matrix:
\begin{align}
{ \rho({\vec r},{\vec r'})}= \sum_a^{N_\psi} f(\varepsilon_a,\mu,T)  \psi_a ({\vec r})  \psi_a^{*}({\vec r\,'})  &
\\\nonumber
+  \sum_b^{N_\chi}  \tilde X_b({\vec r}) \tilde X_b^{*}({\vec r\,'}) &
\end{align}
or  the mixed identity operator:
\begin{align}
\label{iden}
 {\hat I}= \sum_a^{N_\psi}  \vert \psi_a ({\vec r})\rangle \langle \psi_a ({\vec r}) \vert  +  \sum_b^{N_\chi} \vert \tilde \chi_b({\vec r}) \rangle \langle \tilde \chi_b({\vec r}) \vert 
\end{align} 
in expressions for most other observables. Proof of Eq. \ref{iden} is shown in supplemental materials.  In step 6, the chemical potential is found by iteratively solving $N_e=\int d{\vec r}^{\,3} {\rho({\vec r})} \equiv N_e$, were $N_e$ is the number of electrons in the simulation volume \cite{Cytter:2018aa}. When solving for updated $\mu$, as well as the electronic entropy, only the coefficients of the Chebyshev expansion need to be recalculated, using the Cebyshev moments \cite{Cytter:2018aa}. If $N_\psi=0$ or $N_\chi=0$ the algorithm reduces to the standard sDFT or dDFT algorithms respectively.

To illustrate the partitioning of the  eigenspectrum, we show (Figure \ref{dos}), the density of states (DOS), $\text{Tr}\{ {\gamma/\pi\over{(E-{\hat H}_{DFT})^2 + \gamma^2}}\}$, and the occupied DOS, $\text{Tr}\{{f(\hat H_{DFT},\mu,T)\gamma/\pi\over{(E-{\hat H}_{DFT})^2 + \gamma^2}}\}$, for diamond structure carbon at an electron temperature of 10 eV. dDFT, sDFT, and mDFT are compared. Simulation parameters are given in supplemental materials. The results from the different methods are nearly indistinguishable until the dDFT DOS reaches its maximum eigenvalue at ~122 eV. The deterministic contribution to the mDFT DOS has been solid-filled (the artificial tail from finite $\gamma$ is cut-off at the maximum deterministic eigenvalue of $\psi s$ at 12.81 eV). 
   \begin{figure}[t]
\begin{center}
\includegraphics[width=1.0\columnwidth] {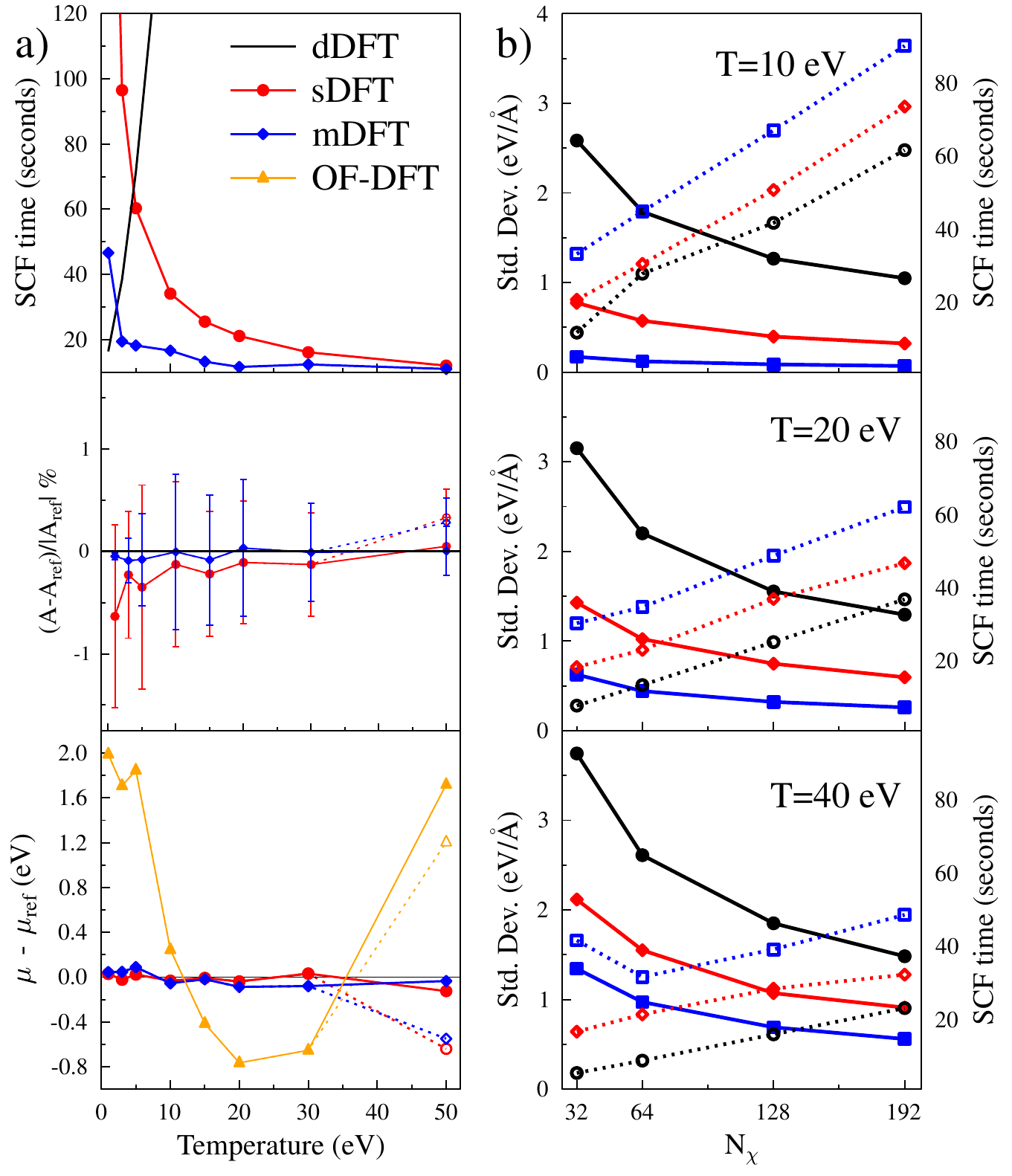}
\caption{(Color online) Single disordered Carbon snapshot 3.51g/cc. (a) \emph{Top}-Comparison of SCF calculation times on 512 threads. See text for parameters. \emph{Middle}- Comparison of relative error in free energy ($A$). \emph{Bottom}-Comparison of electron chemical potential ($\mu$). 0 shown as eye-guide. (b)  Average standard deviation of forces  (solid lines, filled markers) and SCF time on 256 threads (dashed lines, open markers) for different $N_{\chi}$ (x-axis) and $N_{\psi}$. $N_{\psi}=0,128$ and $256$ are shown as black circles, red diamonds, and blue squares respectively.}
\label{DATA}
\end{center}
\end{figure}

The acceleration of the SCF cycle by mDFT is shown in  the top panel of Figure. \ref{DATA}a, where dDFT, sDFT, and mDFT are compared. For the comparison we draw a single snapshot taken from an orbital-free DFT calculation. Any reasonable snapshot will due for this comparison of algorithms. Simulation parameters are given in supplemental materials. We utilize the PBE GGA exchange correlation functional \cite{Perdew:1996aa}. In principle, finite-temperature effects should be accounted for in the exchange correlation functional \cite{PhysRevE.93.063207,PhysRevLett.112.076403,PhysRevB.101.195129,PhysRevLett.115.130402}, however here we are interested only in comparing algorithms so PBE will serve as well as any other GGA. 

All KS-DFT calculations are using our developmental DFT code with two-level band and FFT parallelism \cite{BOTTIN2008329} (using FFTE library \cite{ffte} ), LOBPCG \cite{doi:10.1137/S1064827500366124} for the eigendecomposition, the LibXC package \cite{LEHTOLA20181}, and OpenMPI\cite{10.1007/11752578_29}. sDFT and mDFT calculations are averaged over 20 runs. For sDFT, $N_\chi=120$  yielding $\sim$ 0.5\% relative standard error in the free energy. For dDFT $N_\psi$ was chosen to be high enough to ensure the smallest occupation was reasonably converged, $f(\varepsilon_{N_\psi})<0.0002$. Table \ref{table1} lists the dDFT  $N_\psi$ and the mDFT $N_\psi \, \& \, N_\chi$. By adjusting $N_\psi \, \& \, N_\chi$, while still achieving the same or lower standard deviation in the free-energy, the mDFT SCF time is reduced compared to both the sDFT and dDFT in intermediate temperatures (between 3 and 50 eV). 

The accuracy and precision of the mDFT and sDFT results are demonstrated in the middle and bottom panel of Fig. \ref{DATA}a. The relative difference between the average free-energy sDFT and mDFT results and the reference are shown in the middle panel with error bars $\pm 1$ standard deviation (over the 20 calculations). The reference result is dDFT for 1-30 eV. For 50 eV, the comparison to the highly converged mDFT($N_\psi=4000$ and $N_\chi=400$) is shown by solid line and markers, while comparison to an unconverged dDFT, $f(\varepsilon_{N_\psi})=0.0018$ at $N_\psi=4200$, is shown with open line and markers. The reference values are shown in Table \ref{table1}. Small accuracy differences seen in sDFT (~0.25\%) are due to nonlinear effects (DFT Hamiltonian dependence on density), however the accuracy converges rapidly, as $1/{N_\chi}$, compared to precision, ($1/\sqrt{N_\chi}$)\cite{doi:10.1002/wcms.1412}. These nonlinear effects on accuracy are reduced for mDFT. The bottom panel of Fig. \ref{DATA}a plots the difference of the averages and reference (also listed in Table \ref{table1}) electronic chemical potentials from the same calculations. In addition to KS-DFT calculations, we also compare an orbital-free DFT\cite{Perrot:1979aa,Perdew:1981aa,Lambert:2006aa}. These comparisons verify that the different KS-DFTs converge to the same results, which we already expect analytically.

 \begin{table}[t]
 \begin{tabular}{|c c c c c|} 
  \hline
T (eV) & $\text{mDFT} \atop N_\psi : N_\chi$  &  $\text{dDFT} \atop N_\psi$  & $\text{dDFT} \atop A(eV)$  &  $\text{dDFT} \atop \mu(eV)$ \\ [0.5ex] 
 \hline\hline
 1 & 136 : 4 & 192 &  -123.867& 13.007  \\ 
 \hline
 3 &  128 : 8 & 336 & -126.556 & 12.808  \\
 \hline
 5 & 112 : 8 & 512 & -131.579  & 12.250  \\
 \hline
 10 & 112 : 8 & 1024 & -152.934  & 9.080  \\
 \hline
 15 & 96 : 16 &1600 & --184.592 & 3.650  \\ [1ex] 
 \hline
  20 & 80 : 32 & 2400 & -224.401  & -3.533  \\ [1ex] 
 \hline
  30 &  64 : 40 & 4200 & -322.955& -21.917  \\ [1ex] 
 \hline
   50 & 32:80 & 4200* & -572.859 & -69.619 \\ [1ex] 
   \hline
   \end{tabular}
 \begin{tabular}{|c c c c|} 
 \hline
 T (eV) & $\text{mDFT} \atop N_\psi : N_\chi$   & $\text{mDFT} \atop A(eV)$  &  $\text{mDFT} \atop \mu(eV)$ \\ [0.5ex] 
 \hline\hline
 50 &   4000:400   & -574.466 & -70.134   \\ [1ex] 
 \hline
\end{tabular}
\caption{System is 3.51 g/cc disordered carbon. Number of orbitals and reference values for free energy per atom ($A$) and electron chemical potential ($\mu$), corresponding to Figure \ref{DATA}a.  *$N_{\psi}$ insufficient for converging calculation. Lower table shows the highly converged mDFT used as alternative reference. }
\label{table1}
\end{table}
 
 As previously mentioned, convergence of the electron density and concurrently the nuclear forces requires significantly more stochastic orbitals than the volume averaged energies. In Figure \ref{DATA}b, we show for $T=$10, 20, and 40 eV the standard deviation in the force on the nuclei, calculated from 10 repeated calculations of sDFT and mDFT for the same snapshot used in Figure \ref{DATA}a and then averaged over all nuclei and directions. Calculations are the same parameters and system as above. In general, the random force error can be reduced by either accounting for more electron density deterministically (increasing $N_\psi$) or increasing the number of stochastic orbitals $N_\chi$.  For the latter the error goes  $\propto N_\chi^{-0.5}$, while for the former it depends on $\mu$ and $T$. There are diminishing returns for increasing $N_\psi$ when the maximum eigenenergies become significantly larger than $\mu$. $\mu$ is typically becoming more negative for higher temperatures (at a fixed density), while the cost of calculating occupied stochastic orbitals is decreasing. 
 
 This leads to a general ``rule" that as temperature increases then the value of increasing $N_\chi$ compared to $N_\psi$ goes up. Some points show higher than expected SCF times due to more SCF cycles being required. It should be noted that at $~50$ eV the stochastic algorithm becomes so efficient, and the lowest-energy orbitals start to become so depopulated, that calculating even the lowest energy deterministic orbitals loses its advantage. However, if we pushed temperatures to $O(100-1000)$ of eV, where the 1s core electrons of the carbon start to be ionized, we would need to change from a 4-electron pseudopotential \cite{Hartwigsen:1998aa} to an all(6)-electron pseudopotential. mDFT would again become advantageous over sDFT since the core electrons will be highly occupied. Thus, the benefits of mDFT over sDFT should exist until the atoms become fully ionized. Beyond these trends, the details of whether one should target higher $N_\psi$ or $N_\chi$ will also depend on the density, $E_{cut}$, and the system size. Larger systems will naturally favor higher $N_\chi$ to $N_\psi$ ratios, since the stochastic algorithm is absolutely linear scaling, while the deterministic will only be nearly linear-scaling for sufficiently small $N_\psi$.

 %To set a reference on the importance of stochastic force fluctuations in WDM we can estimate the standard deviation on the random force resulting from a Langevin Thermostat, $\sigma_L$, for these conditions. This is given by $\sigma_L^2 = 2 m_C\gamma T/ \delta t$ where $m_C$ is the mass of a carbon atom and $\delta t$ is the time step. $\delta t$ should be much less than time between collisions, $\tau$, which is estimated as the Wigner-Seitz radius ($r_{WS}$) divided by the thermal velocity ($\sqrt{3m_CT}$), \emph{i.e.} $\delta t \approx y\tau $, with $y<<1$ . To prevent overdamping $\gamma$ should be much less than the time-scale of velocity correlation, which we can roughly estimate as $\tau$, \emph{i.e.}  $\gamma \approx x/\tau $ with $x<<1$. This yields a simple estimaton, $\sigma_L \approx T/r_{WS} \sqrt{6x/y}$. For a reference we have added a constant line for $\sigma_L$ ($r_{WS}=1.11$\si{\angstrom}) with a conservative $x/y=0.002$ to the plots in Fig. \ref{forces}. This indicates that the stochastic fluctuation in the force should be largely insignificant compared to thermal fluctuations. 
    \begin{figure}[t]
\begin{center}
\includegraphics[width=1\columnwidth] {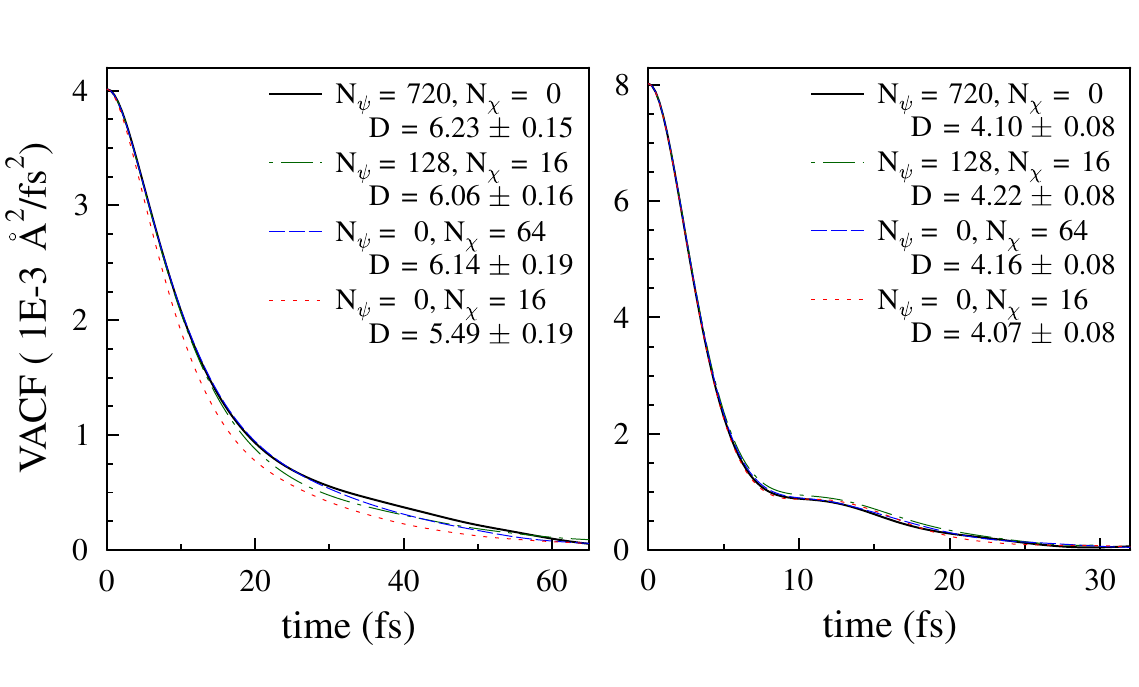}
\caption{(Color online)  Velocity autocorrelation function from dDFT (solidlines), mDFT (dash-dotted line), and two sDFT(dotted and dashed line). 10g/cc 10eV results shown on the right, while 3.51 g/cc 5 eV results are shown on the left. The self-diffusion coefficients ($D$, integral of VACF) are shown in the key with units of $0.001\,cm^2/s$ with the estimated error range (see supplemental for full simulation parameters) .}
\label{vacs}
\end{center}
\end{figure}
 
To verify that the sufficiently small stochastic fluctuations do not affect bulk transport calculations, we have calculated the self-diffusion coefficient \cite{White:2017aa,Meyer:2014aa} of carbon at two WDM conditions, 10 g/cc density with 10 eV temperature, and 3.51 g/cc at 5 eV. Simulation parameters are given in supplemental materials. The coefficients and corresponding velocity autocorrelation functions (VACF) are shown in Figure \ref{vacs}. We compare a fast mDFT calculation with $N_{\psi}=128$ and $N_{\chi}=16$, a ($\sim$2x) slower sDFT calculation with $N_{\chi}=64$, and a fast sDFT calculation with $N_{\chi}=16$  to dDFT. Additional mDFT calculations and comparisons to VASP\cite{Kresse:1999aa} are shown in supplemental materials. All self-diffusion coefficients, $D$, (shown in the key of Fig. \ref{vacs} with units of 0.001 $cm^2/s$)  agree within our estimated error based on the length of the trajectories, with the exception of the sDFT with only $N_{\chi}=16$. Details of VACF generation and error estimation are included in supplemental materials. Molecular dynamics are performed using the iso-kinetic ensemble \cite{Minary03}. These results indicate that the fluctuations due to stochastic forces in (reasonably converged) mDFT/sDFT are not significant contributors to the equilibrium transport properties of WDM. With exception for the $N_\chi=16$ sDFT calculation, the fluctuation of the free energy was not significantly altered by introducing stochasticity. Average and standard deviation in free energies are shown in supplemental materials. These results suggest that, at these very high temperatures, time averaging over equilibrium molecular dynamics reduces the effect of the stochastic force fluctuation, assuming that nonlinear bias is minimized.

In conclusion, we have developed a universal approach to solving the Kohn-Sham DFT self-consistent field calculation which can be efficient at any temperature. The method is sufficient for performing calculation of energies (equation of state) as well as ion transport self-diffusion coefficients, and provides significant acceleration compared to purely deterministic or stochastic DFTs. The performance for viscosity and inter-species diffusion calculations will be examined in the future. Future work will involve generalization of the approach to TD-DFT, calculation of electronic transport coefficients, equation of state and ion transport. This new capability will open a wide range of calculations, previously unachievable at WDM conditions. 
 
 \section*{Acknowledgments}
 We would like to the Prof. Roi Baer and Dr. Edmund Meyer for fruitful discussions. Work supported under the auspices of Science Campaigns 4 and 1 and the Advanced Technical Computing Campaign (ATCC)
 by the US Department of Energy through the Los Alamos National Laboratory. Los Alamos National Laboratory is operated by Triad National Security, LLC, for the National Nuclear Security Administration of U.S. Department of Energy (Contract No. 89233218NCA000001).
 
\bibliographystyle{apsrev4-1}

\end{document}